\documentclass[english]{article}
\usepackage[latin9]{inputenc}
\usepackage{geometry}
\usepackage{amssymb}
\usepackage{graphicx}

\makeatletter

\providecommand{\tabularnewline}{\\}

\usepackage{babel}
\usepackage{placeins}
\usepackage{caption}

\providecommand{\tabularnewline}{\\}

\usepackage{babel}

\makeatother

\usepackage{babel}
\begin{document}

\title{Distribution of Magnetic Dipole Strength -Binning}

\author{Arun Kingan, Mingyang Ma and Larry Zamick\\
 Department of Physics and Astronomy\\
 Rutgers University, Piscataway, New Jersey 08854 }
\maketitle
\begin{abstract}
In previous works we examined the systematics of magnetic dipole transitions in a single j shell. We here extend the study to large space calculations. We consider the nuclei $^{44}$Ti, $^{46}$Ti and $^{48}$Cr. In this work we focus on the B(M1) strength as a function of excitation of energy. The initial state is the lowest J=1$^{+}$T=1 state in a specified
nucleus. The final states are J=0$^{+}$ T=2 , all in one plot, and J=2$^{+}$ T=2 in another. The initial figures have points all over the map although there is a suggestion of an exponential trend. To reduce clutter we perform binning operations in which the summed strength in a given energy interval is represented by a single point. The new
binning curves show more clearly the exponential fall of B(M1)'s with energy.

 \\
 PACs Number: 21.60.Cs 
\end{abstract}

\section{Introduction}

Previously studies of magnetic dipole {[}1,2{]} and Gamow Teller transitions {[}3{]} in the f$_{7/2}$ shell were performed using single j shell programs and results{[}4-7{]}. The nuclei considered were $^{44}$Ti and $^{46}$Ti. In usual scissors mode analyses for even-even nuclei, one starts with a J=0$^{+}$ ground state and the final state is J=1$^{+}${[}8{]}. A notable exception is the work of Beller et al.{[}9{]}. In ref {[}2{]} we started with J=1$^{+}$.From here, there are many more places to
go, e.g. to J=2$^{+}$ T=0,1,2 .One thus gets a better picture of the nature of the collectivity of this state. One striking observation was the fact that there were many strong B(M1) transitions besides the transition back to the ground state.By starting with a J=1$^{+}$ state we are basically connecting to a set of states which constitute the double scissors mode state, as well as to the ground state. 
Selected results for transitions from the lowest J=1$^{+}$state of $^{46}$Ti to various states in a single j shell space (f$_{7/2}$) are show in table 1. We limit the case to final states for which B(M1) is greater than 0.5 \(\mu_{N}^{2}\) . There are many other final
states with smaller values.\\
In a companion paper to this one, in contrast to the single j shell calculations {[}1-3{]}, we performed full f-p shell calculations, focusing on large transition rates and on the spin and orbital content of given transitions{[}10{]}.\\ 
\begin{minipage}[t]{1\columnwidth}
Table 1 :$^{46}$Ti Selected B(M1)'s from the lowest J=1$^{+}$ T=1 state (E{*}= 3.655 MeV) to various (J,T) states, single j (f$_{7/2)}$), MBZE interaction.
\end{minipage}
\begin{center}
\begin{tabular}{|c|c|c|}
\hline 
(J,T) & E{*}(MeV) & B(M1)\(\mu_{N}^{2}\) \tabularnewline
\hline 
\hline 
(0,1) & 0 & 0.560\tabularnewline
\hline 
(0,1) & 4.625 & 2.474\tabularnewline
\hline 
(0,1) & 6.273 & 0.675\tabularnewline
\hline 
(2,1) & 2.496 & 1.202\tabularnewline
\hline 
(2,1) & 3.422 & 1.734\tabularnewline
\hline 
(2,1) & 5.152 & 0.509\tabularnewline
\hline 
(2,1) & 6.158 & 0.648\tabularnewline
\hline 
(2,2) & 8.255 & 1.622\tabularnewline
\hline 
(2,2) & 9.502 & 0.394\tabularnewline
\hline 
\end{tabular}
\end{center}
Much of previous works have focused on the ''scissors mode'' transition from the J=0 ground state to J=1+ states. In our table 1, it is the inverse of the first entry and would have a value \(3 \times 0.560 = 1.68\).
But we see there are much stronger transitions, e.g to a (0,1) state at 4.625 MeV, and to J=2 states with isospins 1 and 2. So what is commonly called the scissors mode transition is not the dominant one. In a companion paper to the current  one, in contrast to the single j shell calculations {[}1-3{]}, we performed full f-p shell calculations, focusing on large transition rates and on the spin and orbital content of given transitions{[}10{]}. The GX1A interaction was used.\\
In this work, we shift from focusing on a few individual strong states to examine the distribution of B(M1) strength as a function of energy. In all tables and figures, B(M1) transitions are given in units of \(\mu_{N}^{2}\) and bare values of the magnetic coupling constants are used.
\section{Comments on the Figures}
In Figure 1, 2, 3, and 4, we show log plots of the B(M1) strengths vs. the energy of the final states for \(^{46}\)Ti and \(^{48}\)Cr. In Fig 1, the final states in \(^{46}\)Ti have quantum numbers J=0 T=2, and in Fig 2, the final states have quantum numbers J=2 T=2. In Figs 3 and 4, we do the same
for \(^{48}\)Cr. The initial state for all 4 figures is the first J=1 T=1 state for \(^{46}\)Ti and \(^{48}\)Cr. Although the points seem at first glance to be all over the place, a closer look shows an exponential decrease of B(M1) strength with energy. one can get a rough fit to Fig 1 with
\(log(B(M1)) = -0.69897 -(E-15)\times 0.20264\) or equivalently, \(B(M1) =0.2 \times 10^{-(E-15)\times 0.20264}\) for E greater than 15 MeV. The important point to make at present is not the precise fit to the calculated points, but rather to make the simple statement that the strength displays
an exponential decrease with energy. To show this more clearly we will, in the next section, introduce a binning process.


\begin{figure}[h]
\centering 
\includegraphics[width=.7\linewidth]{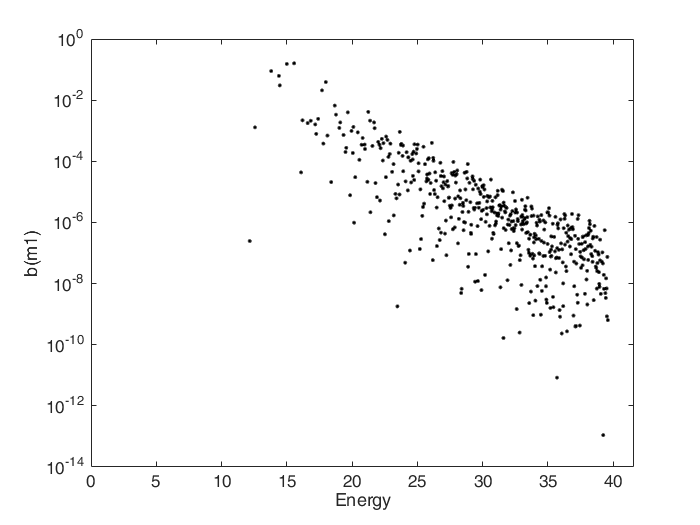}
\caption{$^{46}$Ti B(M1)'s from Lowest J=1 T=1 to Lowest 500 J=0 T=2, Log
Scale }
\label{fig:46Ti10} 
\end{figure}

\begin{figure}
\centering 
\includegraphics[width=.7\linewidth]{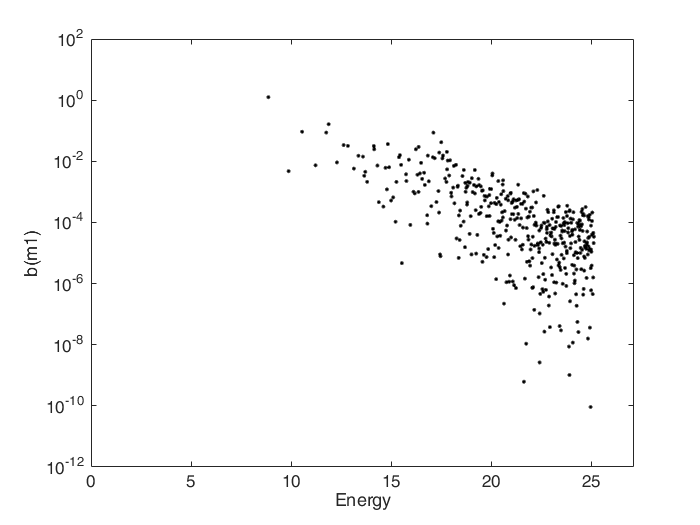}
\caption{$^{46}$Ti B(M1)'s from Lowest J=1 T=1 to Lowest 430 J=2 T=2, Log
Scale }
\label{fig:46Ti12} 
\end{figure}

\begin{figure}
\centering
\includegraphics[width=.7\linewidth]{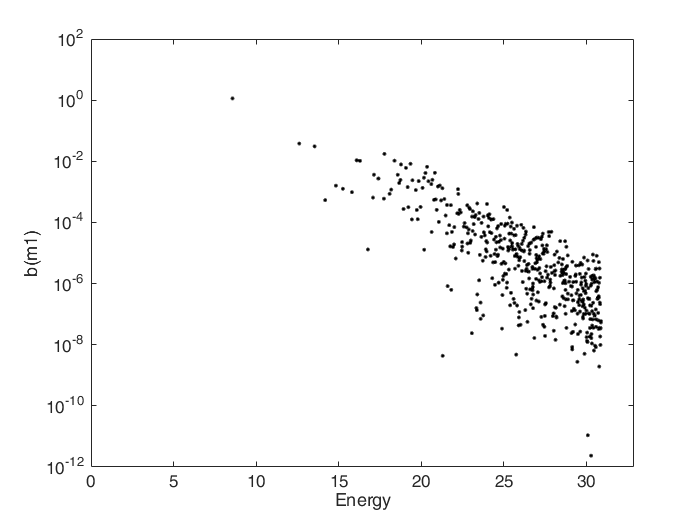}
\caption{$^{48}$Cr B(M1)'s from Lowest J=1 T=1 to Lowest 500 J=0 T=2, Log
Scale}
\label{fig:48Cr10} 
\end{figure}

\begin{figure}
\centering 
\includegraphics[width=.7\linewidth]{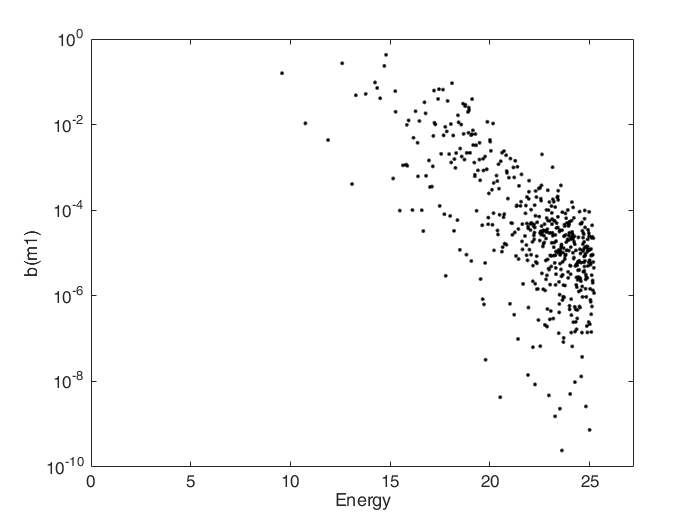}
\caption{$^{48}$Cr B(M1)'s from Lowest J=1 T=1 to Lowest 500 J=2 T=2, Log
Scale}
\label{fig:48Cr12} 
\end{figure}

\clearpage{}\newpage{}

\section{Binning Process}

We see in the above figures that the points are spread all over the place. At the same time we see a rough trend to an exponential decrease of the size of the B(M1)'s. To clarify the matter, we perform what we call a binning process.The results are shown in Fig 5, 6, 7 and 8 of this work. 
When the final states have quantum numbers J=0$^{+}$ T=2, take energy intervals of 0.25 MeV; for J=2$^{+}$ T=2 states we chose the intervals to be 0.2 MeV. For a given energy interval, which we call a bin, we put the sum of all B(M1)'s that fall in that energy range. The log of this sum is represented by a single point on the curve. Results
of this work shown in the 4 figures here should be compared with the corresponding figures in the archived works. Although the new results are far from perfect, there is a great reduction of the clutter. Looking at the results here, we see more clearly an exponential behavior of B(M1 ) for \(^{46}\)Ti as a function of energy (Relative to the initial state
excitation energy). That is the binning B(M1)'s go approximately as \(C e^{-B E}\) where C and B are constants. On a log plot this would be \(log(C) -BE\) i.e. a straight line with a negative slope. We can give a rough fit to Fig I with the formula \(log(B(M1)) = -0.69897 -(E-15)\times 0.20264\) or equivalently, \(B(M1)=0.2 \times 10^{-(E-15)} \times 0.202641)\) for E greater than 15 MeV. In Fig 1 we included 500 states while in Fig 2 430 states. For $^{48}$Cr we used only 100 states (Figs 3 and 4). This is not enough to reach the asymptotic region. We will do improve calculations in the near future. The important point to make at present is not the precise fit to the calculated points but rather to make the simple statement that the strength, after binning,
displays an exponential decrease with energy.

\FloatBarrier

\begin{figure}
\centering
\includegraphics[width=.7\linewidth]{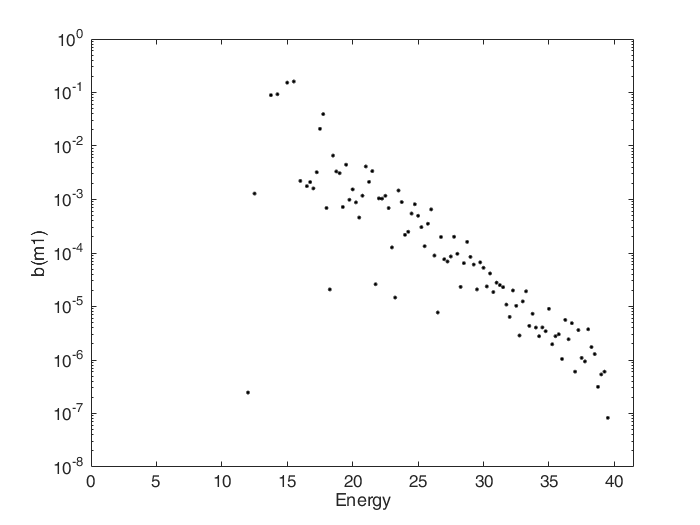}
\caption{$^{46}$Ti B(M1)'s from Lowest J=1 T=1 to Lowest 500 J=0 T=2, Log
Scale, with Binning }
\label{fig:46Ti10bin} 
\end{figure}

\begin{figure}
\centering 
\includegraphics[width=.7\linewidth]{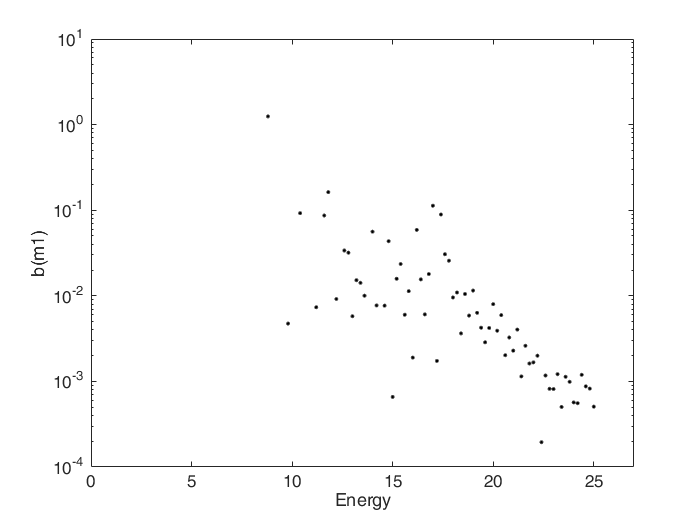}
\caption{$^{46}$Ti B(M1)'s from Lowest J=1 T=1 to Lowest 430 J=2 T=2, Log
Scale, with Binning }
\label{fig:46Ti12bin} 
\end{figure}

\begin{figure}
\centering \includegraphics[width=.7\linewidth]{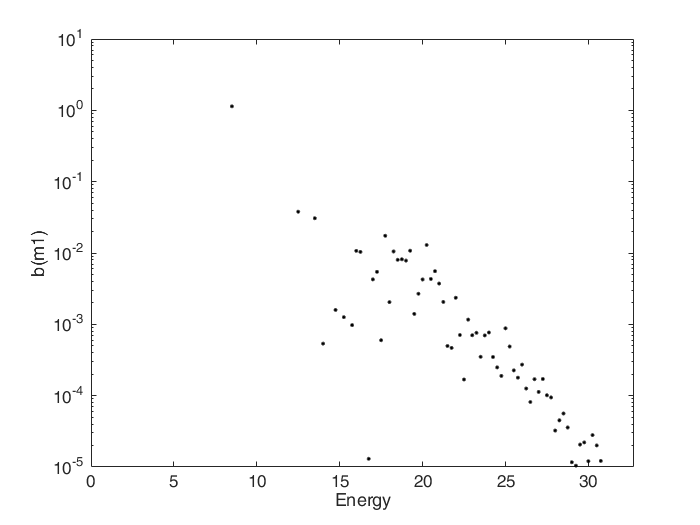}
\caption{$^{48}$Cr B(M1)'s from Lowest J=1 T=1 to Lowest 500 J=0 T=2, Log
Scale, with Binning}
\label{fig:48Cr10bin} 
\end{figure}

\begin{figure}
\centering 
\includegraphics[width=.7\linewidth]{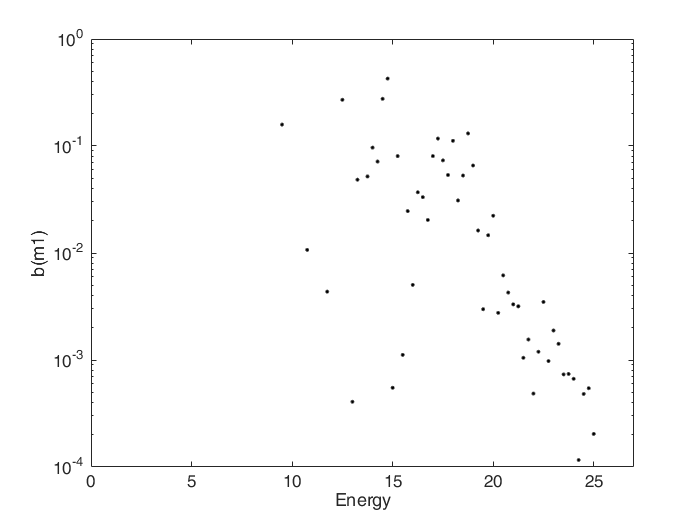}
\caption{$^{48}$Cr B(M1)'s from Lowest J=1 T=1 to Lowest 500 J=2 T=2, Log
Scale, with Binning}
\label{fig:48Cr12bin} 
\end{figure}

\FloatBarrier

\section{Slope analysis}
We add some selected figures in which the best slope line 
is shown. In fig 9, the transitions  in \(^{44}\)Ti are from the lowest J=0 T=0 to J=1 T=0 ;in fig 10, from lowest J=1 T=1 to J=1 T=0; in fig 11 from lowest J=1 T=1 to J=1 T=1 . Note that the final states in all 3 cases are all different and yet the slopes are almost the same. They are \(log(B(M1))=-0.2068E+0.526\), \(log(B(M1))=-0.2091E+0.05206 \), and  \(log(B(M1))=-0.1984E-1.534\) respectively. This slope universality is quite interesting. We do not as of now have a clear explanation. For \(^{46}\)Ti  from J=1 T=1 to J=0 T=1, we  find \(log(B(M1)) =-0.166 E -2.154\).  From J=1 T=1 to J=1 T=2, we find \(log (B(M1)) = -0.152 E -0.861\). In Figures 12 and 13, pertaining to \(^{46}\)Ti, the slopes are \(log(B(M1))= -0.166E-2.094\), \(log(B(M1))= -0.152E-0.861\) respectively for J=1 T=1 to J=0 T=2 and for J=1 T=1
to J=2 T=2. In Fig 14  we consider transitions  from the ground state of \(^{46}\)Ti
J=0 T=1 to J=1 T=1  . We find  \(log(B(M1))=-0.167E-0.124\).We do not have a clear explanation.
\begin{figure}[h]
    \centering
    \includegraphics[width=.7\linewidth]{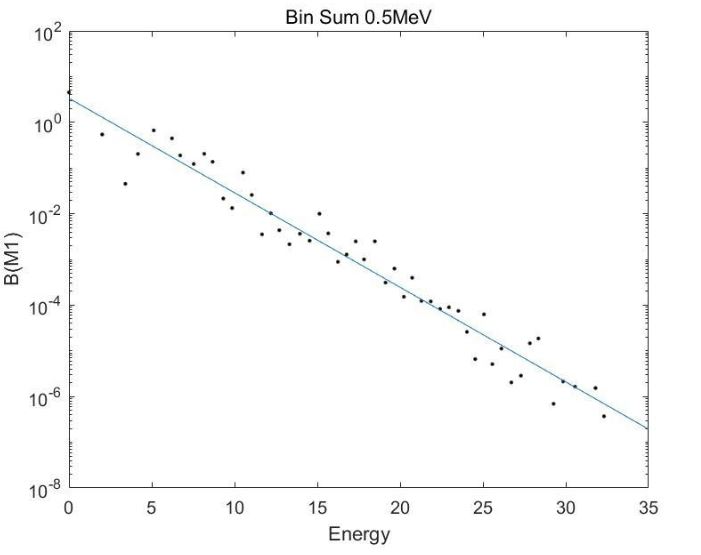}
    \caption{\(^{44}\)Ti B(M1)'s from Lowest J=0 T=0 to Lowest 205 J=1 T=1, Log Scale. with Binning}
    \label{fig:44Ti01bin}
\end{figure}
\begin{figure}
    \centering
    \includegraphics[width=.7\linewidth]{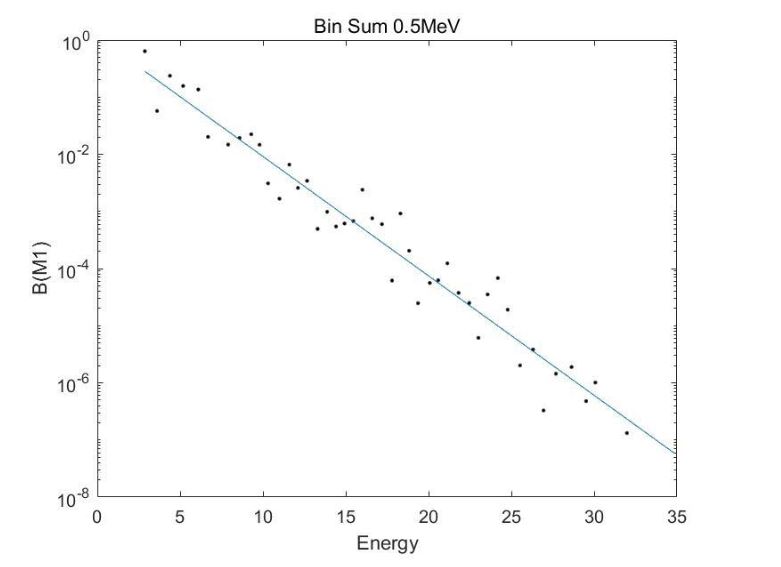}
    \caption{\(^{44}\)Ti B(M1)'s from Lowest J=1 T=1 to Lowest 127 J=1 T=0, Log Scale. with Binning}
    \label{fig:44Ti10bin}
\end{figure}
\begin{figure}
    \centering
    \includegraphics[width=.7\linewidth]{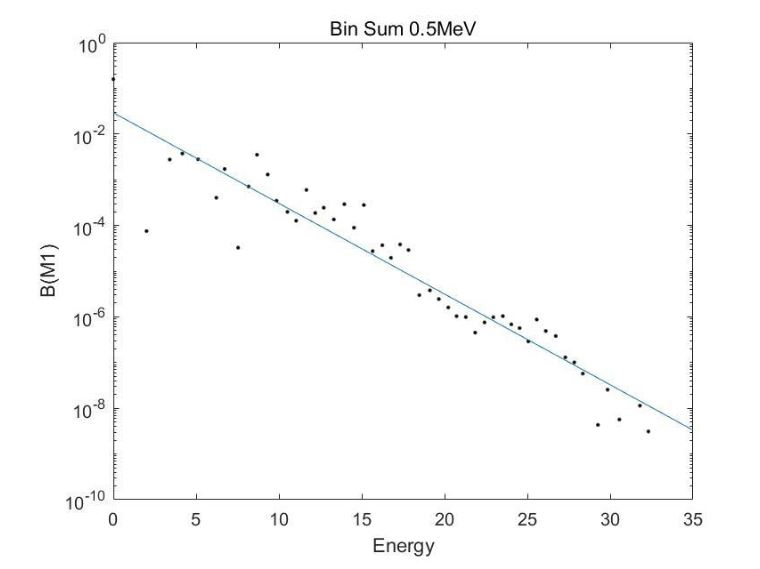}
    \caption{\(^{44}\)Ti B(M1)'s from Lowest J=1 T=1 to Lowest 205 J=1 T=1, Log Scale. with Binning}
    \label{fig:44Ti11bin}
\end{figure}
\begin{figure}
    \centering
    \includegraphics[width=.7\linewidth]{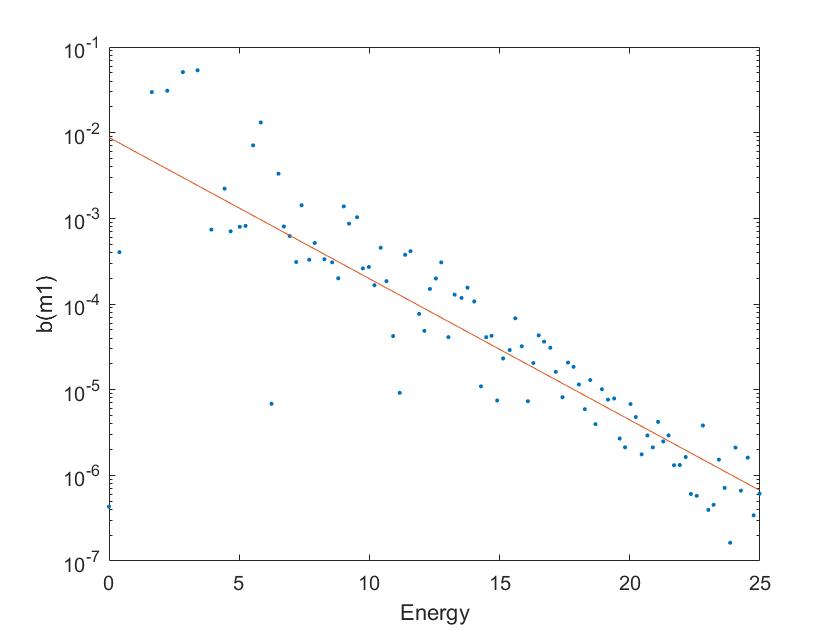}
    \caption{\(^{46}\)Ti B(M1)'s from Lowest J=1 T=1 to Lowest 500 J=0 T=2, Log Scale. with Binning}
    \label{fig:46Ti10bin}
\end{figure}
\begin{figure}
    \centering
    \includegraphics[width=.7\linewidth]{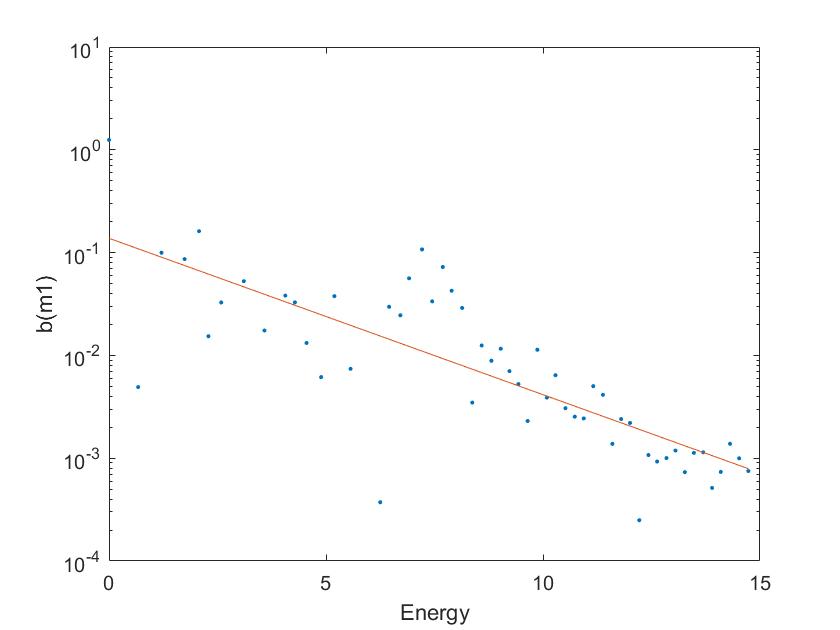}
    \caption{\(^{46}\)Ti B(M1)'s from Lowest J=1 T=1 to Lowest 500 J=2 T=2, Log Scale. with Binning}
    \label{fig:46Ti12bin}
\end{figure}
\begin{figure}
    \centering
    \includegraphics[width=.7\linewidth]{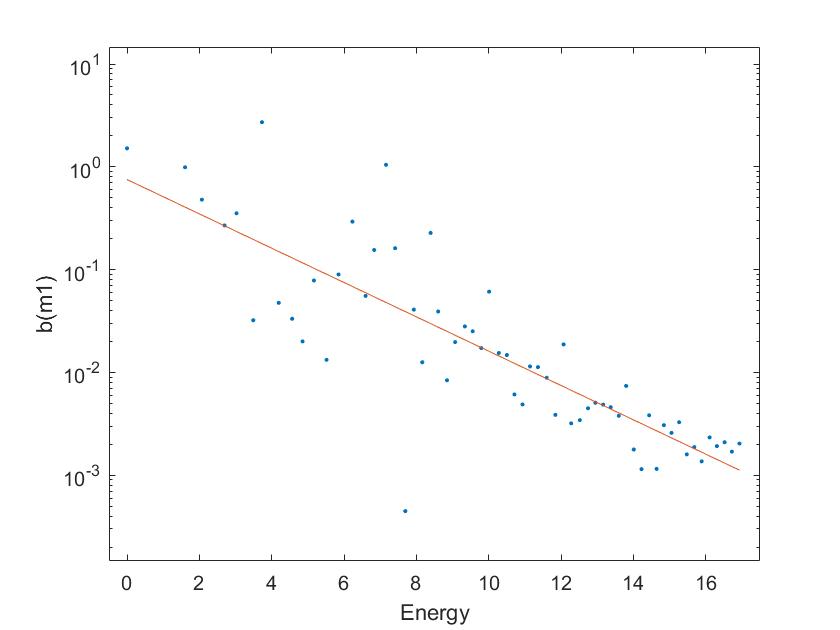}
    \caption{\(^{46}\)Ti B(M1)'s from Lowest J=0 T=1 to Lowest 500 J=1 T=1, Log Scale. with Binning}
    \label{fig:46Ti12bin}
\end{figure}
\FloatBarrier

\section{Closing remarks}

In this work we limited ourselves to the simplest point-how the B(M1) strength varies with energy. We found that initially there is a very wide spread of points. This spread was considerably reduced when we applied the binning process. A near exponential decrease with shift energy was found. Exponential decreases have been noted and calculated
by Schwengner et al. {[}11{]} in processes that are the inverse of electron excitation, e.g. photonuclear reactions. Here the reduced M1 decay probabilities of an excited state are tabulated as a function of the gamma ray energies. There is an exponential fall off with increasing gamma ray energy. This differs from our work here in that we have one fixed initial state (or if we invert things one final state).\\
In the photonuclear case, one considers a cascade of gamma rays from any state to any other state. Related works by Brown et al. {[}12{]}, Schwengner et al. {[}13{]}, Siega {[}14{]} and Karampagia et al. {[}15{]} also deal with these processes. The exponential fall off in these works support our contention that such behavior is widespread.
When we start for the J=0 ground state we reach the scissors mode. When we start form the J=1 state.\\
We construct what is approximately the double scissors mode .This deserves further study. It is reminiscent of earlier  work on double Gamow -Teller states by D.C. Zheng,L. Zamick and N. Auerbach [16,17]. See also H. Sagawa  and T. Uesaka [18], N. Auerbach and B.M. Loc [19] and V. dos S. Ferreira [20].

\section{Acknowledgements}

A.K. received support from the Richard J. Plano Research Award. M.M. was an Aresty student. 
We thank M.Qinonez and .Yu for their contributions to early stages of this work.


\begin{thebibliography}{10}
\bibitem{key-1}M. Harper and L.Zamick, Phys. Rev. C 91,014304 (2015)

\bibitem{key-2} M. Harper and L. Zamick,Phys. Rev. C 91, 054310 (2015)

\bibitem{key-3} R. Garcia and L.Zamick Phys. Rev. C 92, 034322 (20??

\bibitem{key-4}B.F. Bayman, J.D. McCullen and L. Zamick, Phys Rev
Lett $\mathbf{11}$, 215 ($ $1963)

\bibitem{key-5}J.D. McCullen, B.F. Bayman, and L. Zamick, Phys Rev
$\mathbf{134}$,B515 (1964)

\bibitem{key-6}J.N. Ginocchio and J.B. French , Phys. Lett $\mathbf{7}$,
137 (1963)

\bibitem{key-7}A. Escuderos, L.Zamick and B.F. Bayman, Wave functions
in the f$_{7/2}$ shell, for educational purposes and ideas, arXiv:
nucl-th/ 0506050

\bibitem{key-8} K.Heyde, P.von Neumann-Cosel and A. Richter Rev.Mod.Phys.
82, 2365 (2010)

\bibitem{key-9} J. Beller et al. ,Phys. Rev.Lett. 111, 172501 (2013)

\bibitem{key-10}A. Kingan,M.I.Quinonez, X. Yu and L.Zamick,IJMPE
28,1850090 (2018)

\bibitem{key-11}R. Schwengner, S. Frauendorf and A.C. Larson, Phys.
Rev. Lett., 232504 (2013)

\bibitem{key-12}B. Alex Brown and A.C. Larsen Phys. Rev. Lett. 113,
252502 (2014)

\bibitem{key-13}R. Schwengner,S. Frauendorf and B.A. Brown, Phys.Rev.Lett.
118,092502 (2017)

\bibitem{key-14}K. Siega, Phys. Rev. Lett 119 ,052502 (2017)

\bibitem{key-15}S. Karampagia, B.A. Brown and V. Zelevinsky, Phys.
Rev C95, 024322 (2017)
\bibitem{key-16}D.C.Zheng, L.Zamick and N. Auerbach , Phys. Rev.C 40, 936 (1989)

\bibitem{key-1}C.Zheng, L.Zamick and N. Auerbach, Annals of Physics, 197 (1990) 343


\bibitem{key-18}H.Sagawa and T. Uesaka , Phys. Rev. C 94, 064325 (2016)
\bibitem{key-19} N.Auerbach and B.M. Loc ,Phys. Rev. C98, 064301 (2018)
\bibitem{key-20} V. dos S. Ferreira, A.R.Samana,K.Krmpotic and M. Chiapparini,Phys.Rev. C101,044314 (2020)
\end{thebibliography}
\end{document}